\documentclass[11pt]{article}

\usepackage{graphicx,amstext,amssymb}
\usepackage{amsmath}

\begin{document}

\author{S.V. Akkelin \footnote{E-mail: akkelin@bitp.kiev.ua} $^{1}$}
\title{Fluctuations driven  isotropization of the quark-gluon plasma in heavy ion collisions}
 \maketitle

\begin{abstract}

Averaged over ensemble  of initial conditions kinetic transport
equations of  weakly coupled systems of quarks and gluons  are
derived. These equations account for the correlators of
fluctuations of particles and classical gluon fields. The
isotropization of  particle momenta by field fluctuations at the
early prethermal stage of matter evolution in ultrarelativistic
heavy ion collisions  is discussed.  Our results can be useful for
understanding under what  conditions   isotropization of the
quark-gluon plasma in ultrarelativistic heavy ion collisions can
be reached within phenomenologically observed  time scales.

\end{abstract}

\begin{center}

{\small \textit{$^{1}$Bogolyubov Institute for Theoretical
Physics, Metrolohichna str. 14b, 03680 Kiev-143,  Ukraine \\[0pt]
}}



\end{center}

\section{Introduction}

Ideal fluid hydrodynamic models provide a good description  of
Relativistic Heavy Ion Collider  (RHIC) data on single-particle
hadron momentum spectra and elliptic flows \cite{Heinz}. It is
noteworthy that the agreement with data is achieved only if very
rapid equilibration proper times $\tau_{i} < 1$ fm/c are assumed
\cite{time1}, whereas the theoretical estimates based on the
initial conditions obtained from the Color Glass Condensate (CGC)
wave functions of colliding nuclei (see, e.g., Ref. \cite{Iancu})
and perturbative scattering processes yield thermalization times
near $3$ fm/c \cite{time2}. The possible resolution of this puzzle
was proposed in Ref. \cite{Arnold} where it was argued that for
hydrodynamic modeling of the early stage of RHIC collisions
thermalization is not required and isotropization of parton
momenta in the local fluid rest frame (local rest frame of the
energy flow) suffices for applicability of ideal fluid
hydrodynamic models. Then "early thermalization" in heavy ion
collisions is more properly interpreted as evidence of the local
isotropization in momentum space of the nonequilibrium quark-gluon
plasma (QGP), which precedes  the thermalization by acting on a
faster time scale. The gauge fields' instability effects caused by
a particle momentum anisotropy in the local rest frame (it appears
due to the rapid longitudinal expansion), analogous to Weibel
instability in Abelian plasmas \cite{Weibel}, would speed up the
onset of isotropization and subsequent thermalization of weakly
coupled QGP (for review see, e.g., Refs. \cite{Mrow5,Str}) in
relativistic heavy ion collisions because  the unstable modes tend
to make the particle momentum distributions more isotropic. Then a
important question is whether such instabilities could restore
isotropy in momentum space in heavy ion collisions on a  relevant
time scale.

Recently the $3+1$ dimensional  numerical simulations of
nonAbelian plasma instabilities in a stationary anisotropic weakly
coupled plasma ($g \ll 1$ where $g$ is the coupling constant) were
carried out  \cite{Arnold1,Arnold2,Bodeker} using the framework of
the hard-loop (HL) effective theory \cite{HTL,Mrow3,Blaizot}. The
HL effective theory treats the plasma particles as having
arbitrary large momentum; this approximation is justified because
in weak gauge coupling scenarios relevant for the early stage of
relativistic heavy ion collisions $|\textbf{k}|\ll |\textbf{p}|$,
where $|\textbf{k}|$ is the characteristic momentum scale
associated with the unstable gauge fields' modes and
$|\textbf{p}|$ is the typical momentum of plasma particles. This
approach does not account for back reaction of the gauge fields on
the particles because  the gauge fields' effect on the particle
"trajectories" becomes important when the amplitude of the gauge
fields is $A \sim |\textbf{p}|/g$ (in the Coulomb gauge), while in
these simulations $A  \lesssim |\textbf{k}|/g$. Even with this
simplification, the HL dynamics for nonAbelian theories are rather
complex as a result of the nonlinear gauge field self-interactions
that come into play when $A \sim |\textbf{k}|/g$. The simulations
performed for a plasma with moderate momentum anisotropy
\cite{Arnold1} showed that unlike Abelian plasmas (and in contrast
with earlier results of  $1+1$ dimensional simulations
\cite{Rebhan}) the gauge field dynamics changes from exponential
field growth with time to linear growth when the vector potential
amplitude reaches  the nonAbelian scale, $A \sim |\textbf{k}|/g$,
where nonlinear gauge field interactions become important.
Evidently, it could slow down instability-driven particle momentum
isotropization in heavy ion collisions. The same is valid for
$3+1$ dimensional numerical simulations \cite{Arnold2} with
extreme momentum anisotropy and strong initial fields that are
nonperturbatively large, $A \gtrsim |\textbf{k}|/g$. As for the
case of very strong momentum anisotropy and perturbatively weak
initial fields, $A \ll |\textbf{k}|/g$, it was found
\cite{Bodeker} that the exponential growth continues beyond the
nonAbelian bound and extends to higher wave vectors as compared to
the perturbative scenario. The HL effective theory was also
extended to the case of boost-invariant longitudinally expanding
distribution of plasma particles \cite{Rebhan1} and it was
demonstrated in $1+1$ dimensional simulations \cite{Rebhan2} that
chromo-Weibel instabilities grow nearly exponentially in the
square root of proper time \cite{Arnold3}.

Important results have been also obtained beyond the HL
techniques. The nonAbelian collective instabilities were studied
in $3+1$ dimensional numerical simulations \cite{Rom1} of
Yang-Mills equations for unstable matter expanding into the vacuum
after a high energy heavy ion collision. These calculations
account for
 the back reaction of the soft field
modes on the hard modes ("particles").  The calculations performed
for the CGC initial conditions with fluctuations of the fields in
rapidity (violations of boost invariance) demonstrated  that
nonAbelian self-interactions cause the growth of soft modes to
saturate; however,  the isotropization time scales of hard modes
are much shorter if there are large initial fluctuations
\cite{Rom1}. The solution of the full three-dimensional classical
Vlasov transport equations \cite{Dum}
 also goes beyond  the HL approximation. In this kinetic
approach, the mass shell partons released from the wave functions
of relativistic colliding nuclei are treated as "fields", if their
momenta are much below the saturation momentum $Q_{s}$ (given by
the  square root of the color charge density per unit area in the
incoming nuclei), and as "particles", if their  momenta are on the
order of $Q_{s}$ and above. It was  found in numerical simulations
of classical transport equations  that for  fairly strong initial
random fields a very rapid isotropization of the particle momentum
distribution was reached while there was no developed instability
with rapid growth of the fields \cite{Dum}. These results would
indicate that for large initial field fluctuations isotropization
of particle momenta can be reached before the instabilities will
develop.

 These findings probably
indicate  that  thermalization of QGP at RHIC or Large Hadron
Collider (LHC) energies  is mainly an initial state problem and
that fast isotropization in ultrarelativistic heavy ion collisions
can be reached if evolution starts from specific initial
conditions. In this article, we demonstrate using analytical
methods how fluctuations in the initial stage can speed up the
isotropization process. For this purpose we,  based on
collisionless Vlasov transport equations of  QGP
\cite{Elze1,Elze2,Elze3,Mrow1} and using the methods developed for
Abelian plasmas \cite{Kl1,Kl2,Kl3,Ts} (see also Ref. \cite{Litim}
where effective transport equations for nonAbelian plasmas are
derived based on Wong equations \cite{Wong}), derive kinetic
equations of weakly coupled QGP (wQGP) that describe fluctuations
driven isotropization  of the averaged over ensemble of initial
conditions   distribution functions of on-mass-shell "particles"
(quarks, antiquarks  and gluons). The turbulent nonAbelian plasma
instabilities are accounted for in  an approach that was developed
early for the turbulent Coulomb plasmas \cite{Vedenov} (for
review, see Ref. \cite{Ts}).

This article  is organized  as follows. In Sec. 2, we derive
averaged over ensemble  of initial conditions  kinetic transport
equations of wQGP  based on a collisionless approximation of
kinetic equations. The "collision contributions", which naturally
appear in these equations, are expressed through the correlators
of fluctuations.
 In Sec. 3, we
consider  averaged wQGP kinetic equations  below the nonAbelian
scale where the nonlinear gauge fields' self-interactions can be
ignored and we derive the corresponding  collision terms that
describe  diffusion in momentum space leading to the locally
 momentum  isotropical state as well as
generalized Balesku-Lenard "collisions" leading to the (local)
equilibrium state  of the particle phase-space distributions. We
conclude in Sec. 4.

\section{Averaged over ensemble  of initial conditions  kinetic transport equations of wQGP }

Let us start this section with  a brief   review of the transport
equations of QGP in the collisionless Vlasov approximation.   The
collisionless approximation is  applied to kinetic equations that
describe the evolution of  the distribution functions that are
smoothed over physically small volumes\footnote{In a certain
sense, utilization of "smoothed" quantities, and, so, transition
from deterministic to probabilistic description, is unavoidable
for  macroscopic systems because it is impossible to fix
("observe") the micro-state  of the macro-system with absolute
accuracy without destroying-out the macro-state \cite{Krylov}.
Note  that even for exact initial conditions the "smoothing"
appears effectively  for the quantities that are calculated (in
each "event") by means of numerical molecular dynamic models that
are the "solver" of the reversible Hamiltonian dynamic equations.
This is the result of the stochastic errors of the "round-up" of
the numbers and the systematic errors of the method that are
"triggered" by the dynamic chaos that is inherent to complex
Hamiltonian systems.} \cite{Kl1}, and this means neglect of
short-range fluctuations that are responsible for the appearance
of the Boltzmann collision terms in kinetic equations
\cite{Kl1,Kl2} and, so, neglect of large angle ("hard") scattering
of particles due to high transferred momenta at short distances.
This approximation is justified for time scales that are short
compared to the mean time between large angle scatterings of
plasma particles.

The distribution functions of quarks, $Q(x,p)$, antiquarks,
$\overline{Q}(x,p)$, and gluons, $G(x,p)$, are assumed to satisfy
the following collisionless nonAbelian  Vlasov-type transport
equations that describe high momentum modes that are treated as
classical colored particles and soft gluons that are treated as
classical fields (for details see, e.g., Refs.
\cite{Mrow5,Mrow3,Elze3,Mrow1,Mrow2,Mrow4}):
\begin{eqnarray}
p^{\mu}D_{\mu}Q(x,p)+\frac{g}{2}p^{\mu}\{F_{\mu \nu},
\frac{\partial Q(x,p)}{\partial p_{\nu}}\}=0, \label{micro-q} \\
p^{\mu}D_{\mu}\stackrel{-}{Q}(x,p)-\frac{g}{2}p^{\mu}\{F_{\mu
\nu}, \frac{\partial \stackrel{-}{Q}(x,p)}{\partial p_{\nu}}\}=0,
\label{micro-antiq} \\
p^{\mu}\widehat{D}_{\mu}G(x,p)+\frac{g}{2}p^{\mu}\{\widehat{F}_{\mu
\nu}, \frac{\partial G(x,p)}{\partial p_{\nu}}\}=0.
\label{micro-g}
\end{eqnarray}
Here $x=(t,\textbf{r})$, $p=(p_{0},\textbf{p})$,
\begin{eqnarray}
D_{\mu}= I\partial_{\mu}-ig [A_{\mu}(x),...], \label{def-q} \\
\widehat{D}_{\mu}= \widehat{I}\partial_{\mu}-ig
[\widehat{A}_{\mu}(x),...], \label{def-g}
\end{eqnarray}
with $A_{\mu}$ and $\widehat{A}_{\mu}$ being four-potentials,
$F_{\mu \nu}$ and $\widehat{F}_{\mu \nu}$ being  field strength
tensors in the fundamental and adjoint representations,
respectively,
\begin{eqnarray}
A_{\mu}= A^{a}_{\mu}t_{a}, \label{def-g1} \\
\widehat{A}_{\mu}= A^{a}_{\mu}T_{a}, \label{def-g2} \\
F_{\mu \nu}= \partial_{\mu} A_{\nu}- \partial_{\nu} A_{\mu} -
ig[A_{\mu},A_{\nu}], \label{def-f1} \\
\widehat{F}_{\mu \nu}= \partial_{\mu} \widehat{A}_{\nu}-
\partial_{\nu} \widehat{A}_{\mu} - ig[\widehat{A}_{\mu},\widehat{A}_{\nu}], \label{def-f2}
\end{eqnarray}
where $t_a$ and $T_{a}$ are $SU(N_{c})$ group generators in the
fundamental ($N_{c}\times N_{c}$  matrices) and adjoint
[$(N^{2}_{c}-1)\times(N^{2}_{c}-1)$ matrices] representations,
respectively, $a=1,...,(N_{c}^{2}-1)$,  and  the Einstein
summation convention for repeated indices is utilized. The
$[...,...]$ denotes the commutator, and $\{...,...\}$ denotes the
anticommutator.

 Equations (\ref{micro-q})-(\ref{micro-g}) are supplemented by the Yang-Mills equation
 \begin{eqnarray}
D_{\mu}F^{\mu \nu}(x)=-j^{\nu}(x), \label{yang-m}
\end{eqnarray}
 where  the color current density $j_{\mu}$ is expressed in the fundamental representation,
$j_{\mu}=j_{\mu}^{a}t^{a}$, as
\begin{eqnarray}
j^{\mu}= \frac{g}{2} \int d^{4}p \frac{p^{\mu}}{p_{0}}
(Q(x,p)-\overline{Q}(x,p)-\frac{1}{N_{c}}Tr[Q(x,p)-\overline{Q}(x,p)]+2t_{a}Tr[T_{a}G(x,p)]).
\label{def-current}
\end{eqnarray}

Then, to derive   kinetic equations for mean (statistically
averaged) values  one needs to perform ensemble average, $\langle
... \rangle$, of  Eqs. (\ref{micro-q}-(\ref{micro-g}) and
(\ref{yang-m})  over ensemble of initial conditions.  The ensemble
average allows split quark, antiquark, and  gluon phase-space
densities and gluonic classical fields into their mean part and a
fluctuating part, for example,
 \begin{eqnarray}
 A^{\mu}=\langle A^{\mu} \rangle + \delta A^{\mu} .
\label{def-split}
\end{eqnarray}
 The mean value of the statistical fluctuations vanishes by definition, $\langle \delta A^{\mu} \rangle=0$.

  The quark and antiquark phase-space densities $Q$,
$\overline{Q}$ are $N_{c}\times N_{c}$ hermitian  matrices in
color space and  have singlet  and multiplet  parts in the
fundamental representation of the $SU(N_{c})$ gauge group, and the
gluon phase-space density $G$ is an
$(N^{2}_{c}-1)\times(N^{2}_{c}-1)$ hermitian matrix in color space
with singlet and multiplet  parts in the adjoint representation.
We  assume  that the statistically averaged  value of the partonic
phase-space density
 is the singlet (and, so, is locally colorless)  and the disturbance (fluctuation) of
 the phase-space density is the multiplet of the $SU(N_{c})$ gauge
 group. Then
\begin{eqnarray}
Q= \langle Q \rangle + \delta Q = I f_{q} + \delta f^{a}_{q}t_{a},
\label{def-q1} \\
\overline{Q}= \langle \overline{Q} \rangle + \delta \overline{Q} =
I f_{\overline{q}} + \delta f^{a}_{\overline{q}}t_{a}, \label{def-q2} \\
G= \langle G \rangle + \delta G = I f_{g} + \delta f^{a}_{g}T_{a}.
\label{def-g3}
\end{eqnarray}
 We assume    that averaged values of the quark and antiquark  phase-space densities coincide,
 $f_{q}=f_{\overline{q}}$, and that the
statistically
 averaged local value of the classical  gluon field, $A^{a}_{\mu}$, and the  gluon field
 strength, $F_{\mu \nu}^{a}$,
 are equal to zero, $\langle A_{\mu}^{a} \rangle = \langle F_{\mu \nu}^{a} \rangle =
 0$, and  then   $A_{\mu}^{a}= \delta A_{\mu}^{a}$ and
\begin{eqnarray}
 F_{\mu \nu}^{a} = \delta F_{\mu
 \nu}^{a}=\partial_{\mu} \delta A_{\nu}^{a}- \partial_{\nu} \delta A_{\mu}^{a}
 + g \delta A_{\mu}^{c} \delta A_{\nu}^{d}f^{c d a}.
\label{def-delta-f}
\end{eqnarray}
Here we take into account that
\begin{eqnarray}
[t^{a},t^{b}] = i f^{abc} t^{c} , \quad [T^{a},T^{b}] =i f^{abc}
T^{c} , \label{condit-1}
\end{eqnarray}
where $f^{abc}$ are the antisymmetric $SU(N_{c})$  structure
constants.  Keeping  quadratic
 terms in fluctuations, $g \delta A_{\mu}^{c}\delta A_{\nu}^{d}f^{c d a}$, is necessary because  one cannot ignore the
 nonAbelian gauge fields'  self-interactions when corresponding
 amplitudes of fluctuations reach the nonAbelian scale where
  $( \partial_{\mu} \delta A_{\nu}^{a}- \partial_{\nu} \delta A_{\mu}^{a} ) \sim g \delta A_{\mu}^{c} \delta A_{\nu}^{d}f^{c d a}$.
 Note here that the statistical average
 of the gluon field strength (stress tensor),  $\langle F_{\mu \nu}\rangle$, is not only
 given by $F_{\mu \nu}(\langle A \rangle)$ due to quadratic
 terms in the fluctuations  contained in $F_{\mu \nu}$;
 therefore condition $\langle F_{\mu \nu} \rangle = 0$
 implies
\begin{eqnarray}
 \langle \delta
A_{\mu}^{a} \delta A_{\nu}^{b} \rangle \sim \delta^{ab},
\label{def-f1n}
\end{eqnarray}
and, so,  we conclude that the vanishing averaged local value of
the classical gluon field strength means that fluctuations of
different color components of classical gluon fields are
statistically independent.

Performing the ensemble average of  Eqs.
(\ref{micro-q})-(\ref{micro-g}) and  (\ref{yang-m})  gives
\begin{eqnarray}
p^{\mu}\langle D_{\mu}  Q \rangle  + \frac{g}{2}p^{\mu}\langle
\{\delta F_{\mu \nu}, \frac{\partial \delta Q}{\partial p_{\nu}}
\}\rangle  = 0,
 \label{av-q}\\
 p^{\mu}\langle D_{\mu}  \overline{Q }\rangle  -
\frac{g}{2}p^{\mu}\langle \{\delta F_{\mu \nu}, \frac{\partial
\delta \overline{Q}}{\partial p_{\nu}} \}\rangle = 0,
 \label{av-antiq}\\
 p^{\mu}\langle \widehat{D}_{\mu}  G \rangle  +
\frac{g}{2}p^{\mu}\langle \{\delta \widehat{F}_{\mu \nu},
\frac{\partial \delta G}{\partial p_{\nu}} \}\rangle  = 0,
 \label{av-g}
\end{eqnarray}
and
\begin{eqnarray}
\langle D_{\mu} \delta F^{\mu \nu}\rangle = 0.
 \label{av-yang-m}
\end{eqnarray}

 For further convenience let us introduce  mass-shell
distribution functions $f_{i}(x,\mathbf{p})$ and $\delta
f_{i}(x,\mathbf{p})$ (index $i$ here means $q$, $\overline{q}$, or
$g$) depending on four-position $x=(t,\mathbf{r})$ and
three-momentum $\textbf{p}$, e.g.,
\begin{eqnarray}
f_{q}(x,p)=2 p_{0} \Theta  (p_{0}) \delta
(p^{2}-m^{2}_{q})f_{q}(x,\textbf{p})= \delta
(p_{0}-\sqrt{\textbf{p}^{2}+m_{q}^{2}}) f_{q}(x,\textbf{p}) .
 \label{f-new}
\end{eqnarray}
 For the adopted normalization
\begin{eqnarray}
\int d^{4}p f_{q}(x,p)=\int d^{3}\mathbf{p}
f_{q}(x,\textbf{p})=n_{q}(x) ,
 \label{normal-new}
\end{eqnarray}
 where $n_{q}(x)$ is quark number density.

The distribution functions of quarks, antiquarks,  and gluons have
no simple probabilistic interpretation due to  the gauge
dependence since a color of a particle (e.g., a quark) can be
changed by means of a gauge transformation. Only the traces of the
distribution functions are gauge independent and therefore they
have the
 probabilistic interpretation (see, e.g., Ref. \cite{Mrow2}). To get
equations that govern the evolution of the colorless mean values
and, so, present the average (most probable) evolution of the
ensemble of systems,  one can calculate \textit{Trace} of Eqs.
(\ref{av-q})-(\ref{av-g}).\footnote{The \textit{Trace} of Eq.
(\ref{av-yang-m}) leads to trivial identity $0=0$.} Because
\textit{Trace} of the color multiplet part is equal to zero, the
\textit{Trace} explicitly reveals the evolution of the colorless
(singlet) quantities, and the statistically averaged collision
term does not change the color neutrality. Performing integration
in Eqs. (\ref{av-q})-(\ref{av-g}) over $p_{0}$ and taking into
account that $Tr(t^{a})=Tr(T^{a})=0$  and that $Tr(AB)=Tr(BA)$ for
any quadratic matrices of the same order, we get\footnote{Starting
from here the derivatives over $p_{0}$ anywhere throughout the
article are identically zero and $p_{0}\equiv
E_{p}=\sqrt{m^{2}+\textbf{p}^{2}}$.}
\begin{eqnarray}
 p^{\mu}\partial_{\mu} f_{q}(x,\textbf{p}) = - \frac{g}{N_{c}}p^{\mu}
Tr  ( \langle  \delta F_{\mu \nu}(x) \frac{\partial \delta Q
(x,\textbf{p})}{\partial p_{\nu}}  \rangle  ) ,
 \label{tr-av-q}\\
 p^{\mu}\partial_{\mu} f_{\overline{q}}(x,\textbf{p}) =
\frac{g}{N_{c}}p^{\mu}Tr (\langle  \delta F_{\mu \nu}(x)
\frac{\partial \delta \overline{Q}(x,\textbf{p})}{\partial
p_{\nu}}  \rangle ) ,
 \label{tr-av-antiq}\\
 p^{\mu}\partial_{\mu} f_{g}(x,\textbf{p}) =  -
\frac{g}{ (N^{2}_{c}-1)} p^{\mu}Tr (\langle  \delta
\widehat{F}_{\mu \nu}(x) \frac{\partial \delta G
(x,\textbf{p})}{\partial p_{\nu}} \rangle ) .
 \label{tr-av-g}
\end{eqnarray}
 Equations (\ref{tr-av-q})-(\ref{tr-av-g}) describe the evolution of the  averaged over ensemble
of initial conditions phase-space densities of quarks, antiquarks
and gluons  with collision terms that are determined by the
correlators (statistically averaged products) of fluctuations.
These "collision contributions" contain entire physics of "soft"
(long distance) processes, e.g., isotropization in momentum space
due to the chromodynamic Weibel instabilities and establishment of
(local) equilibrium (thermalization). Further analysis is limited
to the quark collision term
\begin{eqnarray}
I_{q}(x,\textbf{p}) \equiv  - \frac{g}{N_{c}}p^{\mu} Tr  ( \langle
\delta F_{\mu \nu}(x) \frac{\partial \delta Q
(x,\textbf{p})}{\partial p_{\nu}}  \rangle ), \label{coll-qq-def}
\end{eqnarray}
 but expansion
of the results to antiquark,  $I_{\overline{q}}$, and gluon,
$I_{g}$,  collision terms is straightforward. The detailed
calculation of $I_{q}$ is carried out in the next section for
amplitudes of fluctuations below the nonAbelian scale; here we
only demonstrate why the collision term, $I_{q}$, can lead to
momentum space diffusion of particle phase-space density, $f_{q}$,
and, so, to local isotropization in momentum space. To demonstrate
it, let us rewrite $I_{q}$ in a slightly different  form,
\begin{eqnarray}
I_{q}(x,\textbf{p}) =   - \frac{g}{N_{c}} \frac{\partial
}{\partial p_{\nu}} Tr  ( \langle p^{\mu} \delta F_{\mu \nu}(x)
\delta Q(x,\textbf{p}) \rangle ) + \frac{g}{N_{c}} Tr  ( \langle
\delta F_{\mu }^{\mu}(x) \frac{\partial \delta Q
(x,\textbf{p})}{\partial p_{\nu}}  \rangle ) .
\label{coll-qq-def-1}
\end{eqnarray}
Note here that the second term on the right-hand side of the above
expression disappears in the Abelian approximation when, in
particular, the stress tensor $\delta F_{\mu \nu}$ is approximated
by $\partial_{\mu}\delta A_{\nu} - \partial_{\nu}\delta A_{\mu}$.
 Then, subtracting  Eq. (\ref{av-q}) from
 Eq. (\ref{micro-q}), performing integration over $p_{0}$, and accounting
for the leading order in fluctuations,\footnote{We keep here
non-linear terms in the covariant derivative and in  the gluon
field stress tensor  because  the corresponding nonAbelian  terms
cannot be neglected if the amplitudes of fluctuations reach the
nonAbelian scale.} we get
\begin{eqnarray}
p^{\mu}D_{\mu} \delta Q (x,\textbf{p}) =  - g p^{\mu} \delta
F_{\mu \nu}(x) \frac{\partial \langle Q \rangle
(x,\textbf{p})}{\partial p_{\nu}}- ig p^{\mu} \langle[\delta
A_{\mu}(x),\delta Q]\rangle .
 \label{flukt-nonab-q}
\end{eqnarray}
 Here we take into account that $[\delta A_{\mu}(x),\langle Q
\rangle]=0$ and that $(1/2) \{\delta F_{\mu \nu}, \frac{\partial
\langle Q \rangle}{\partial p_{\nu}} \}=\delta F_{\mu \nu}
\frac{\partial \langle Q \rangle}{\partial p_{\nu}}$.
 The   formal solution of Eq. (\ref{flukt-nonab-q}) has the
following form:
\begin{eqnarray}
\delta Q(x,\textbf{p}) = -g \int d^{4}y G_{p}(x-y)U(x,y)p^{\mu}
\delta F_{\mu \nu}(y)U(y,x)\frac{\partial \langle Q \rangle
(y,\textbf{p})}{\partial p_{\nu}} + (\textit{other terms}).
 \label{flukt-nonab-q-sol}
\end{eqnarray}
Here
\begin{eqnarray}
p^{\mu}\partial_{\mu}G_{p}(x)=\delta^{(4)}(x) , \label{green1} \\
G_{p}(x)=E_{p}^{-1}\Theta (t)
\delta^{(3)}(\mathbf{r}-(\mathbf{p}/E_{p})t), \label{green2}
\end{eqnarray}
and $U(x,y)$ is the gauge parallel transporter (see, e.g., Refs.
\cite{Mrow3,Blaizot,Elze3}) along the straight line $\gamma$ going
from $y$ to $x$,
\begin{eqnarray}
U(x,y) = {\cal P}\: {\rm exp} \Big[ - ig \int_\gamma dz_{\mu} \:
A^{\mu}(z) \Big] \label{transp},
\end{eqnarray}
where ${\cal P}$ denotes path  ordering. Note that in the Abelian
approximation, when the terms that are not of the leading order in
$g$ are neglected,  the transporter $U(x,y)$ is approximated by
unity. Then, substituting Eq. (\ref{flukt-nonab-q-sol}) for
$\delta Q(x,\textbf{p})$ in the first term of Eq.
(\ref{coll-qq-def-1}) we get
\begin{eqnarray}
I_{q}(x,\textbf{p}) =  \frac{g^2}{N_{c}} \frac{\partial}{\partial
p_{\nu}}\int d^{4}y G_{p}(x-y) {\cal T}_{\nu
\beta}(x,y)\frac{\partial \langle Q \rangle
(y,\textbf{p})}{\partial p_{\beta}} + (\textit{other terms}),
\label{coll-qq-nonab-1}
\end{eqnarray}
where
\begin{eqnarray}
{\cal T}_{\nu \beta}(x,y) \equiv   Tr  ( \langle p^{\mu}\delta
F_{\mu \nu}(x)U(x,y) p^{\alpha} \delta F_{\alpha \beta}(y)U(y,x)
\rangle ). \label{coll-qq-nonab-2}
\end{eqnarray}
Now, if we  assume that the
 correlation lengths in $(x-y)$ of ${\cal T}_{\nu \beta}$  are far less
 than the time and length scales on which  distribution function $\langle Q \rangle$
 varies, we get
 \begin{eqnarray}
I_{q}(x,\textbf{p}) = \frac{\partial}{\partial p_{\nu}} D_{ \nu
\beta} (x,\textbf{p})\frac{\partial f_{q}(x,\textbf{p})}{\partial
p_{\beta}} + (\textit{other terms}), \label{coll-qq-nonab-3}
\end{eqnarray}
where
 \begin{eqnarray}
 D_{\nu \beta } (x,\textbf{p}) \equiv
\frac{g^2}{N_{c}} \int d^{4}y G_{p}(x-y) {\cal T}_{\nu
\beta}(x,y), \label{coll-qq-nonab-4}
\end{eqnarray}
and  we take into account that $\langle Q \rangle = f_{q}I$.
Thereby,  $I_{q}$ contains  a term that  describes diffusion in
momentum space of particle phase-space density and, therefore,
$I_{q}$ (as well as  $I_{\overline{q}}$, $I_{g}$) can be related
with isotropization processes in  relativistic heavy ion
collisions.

\section{Kinetics of wQGP below the nonAbelian scale}
 The main difficulty in calculating the collision
terms is caused by the gauge fields' self-interactions that take
place because of the nonAbelian nature of the QCD. In this section
we discuss
 simple and theoretically
clean situations when  amplitudes of fluctuations  are below the
nonAbelian scale. Then one  can neglect self-interactions and
substitute $F_{\mu \nu} \rightarrow {\cal F}_{\mu \nu}\equiv
\partial_{\mu}\delta A_{\nu} - \partial_{\nu}\delta A_{\mu}$,
$D_{\mu} \rightarrow  \partial_{\mu}$.

To obtain the explicit expressions for the correlators of
fluctuations one needs to derive evolutional equations for the
fluctuations. Then, subtracting averaged Eqs.
(\ref{av-q})-(\ref{av-yang-m}) from Eqs.
(\ref{micro-q})-(\ref{micro-g}) and (\ref{yang-m}), performing
integration over $p_{0}$, and neglecting the terms that are not of
the leading order in $g$, we get
\begin{eqnarray}
p^{\mu}\partial_{\mu} \delta Q (x,\textbf{p}) =  - g p^{\mu}
\delta {\cal F}_{\mu \nu}(x) \frac{\partial \langle Q \rangle
(x,\textbf{p})}{\partial p_{\nu}} ,
 \label{flukt-q}\\
p^{\mu}\partial_{\mu} \delta \overline{Q}(x,\textbf{p})  =  g
p^{\mu} \delta {\cal F}_{\mu \nu}(x) \frac{\partial \langle
\overline{Q} \rangle (x,\textbf{p})}{\partial p_{\nu}} ,
 \label{flukt-antiq}\\
 p^{\mu}\partial_{\mu} \delta G (x,\textbf{p}) =  -
g p^{\mu} \delta \widehat{{\cal F}}_{\mu \nu}(x) \frac{\partial
\langle G \rangle (x,\textbf{p})}{\partial p_{\nu}} ,
 \label{flukt-g}
\end{eqnarray}
and
\begin{eqnarray}
\partial_{\mu} \delta {\cal F}^{\mu \nu }(x) = - \delta j^{\nu}(x). \label{yang-m-fl}
\end{eqnarray}
 The color current fluctuation, $\delta
j^{\mu}(x)$,  is expressed through fluctuations of the quark,
antiquark  and gluon phase-space densities and in the fundamental
representation reads
\begin{eqnarray}
\delta j^{\mu}(x)= g \int d^{3}\textbf{p} \frac{p^{\mu}}{2E_{p}}
(\delta
Q(x,\textbf{p})-\delta \overline{Q}(x,\textbf{p})+2t_{a}Tr[T_{a}\delta G(x,\textbf{p})])= \nonumber \\
g \int d^{3}\textbf{p}\frac{p^{\mu}}{2E_{p}} t_{a}( \delta
f_{q}^{a}(x,\textbf{p})- \delta
f_{\overline{q}}^{a}(x,\textbf{p})+2 N_{c}\delta
f_{g}^{a}(x,\textbf{p})) . \label{def-current-fluct}
\end{eqnarray}
 The solution of  Eqs. (\ref{flukt-q})-(\ref{flukt-g}) has the form
\begin{eqnarray}
 \delta f_{q}^{a} =   \delta f_{q}^{a(s)} + \delta f_{q}^{a(i)} ,
 \label{flukt-q-1}\\
\delta f_{\overline{q}}^{a} =   \delta f_{\overline{q}}^{a(s)} +
\delta f_{\overline{q}}^{a(i)} ,
 \label{flukt-antiq-1}\\
 \delta f_{g}^{a}  =   \delta f_{g}^{a(s)} + \delta f_{g}^{a(i)} ,
 \label{flukt-g-1}
\end{eqnarray}
where $\delta f_{q}^{a(s)}$, $\delta f_{\overline{q}}^{a(s)}$, and
$ \delta f_{g}^{a(s)}$  are associated with the spontaneous
fluctuations of the freely moving partons
\begin{eqnarray}
p^{\mu}\partial_{\mu}\delta f_{q}^{a(s)}(x,\textbf{p})
=p^{\mu}\partial_{\mu}\delta
f_{\overline{q}}^{a(s)}(x,\textbf{p})=p^{\mu}\partial_{\mu}\delta
f_{g}^{a(s)}(x,\textbf{p})=0 , \label{flukt-free}
\end{eqnarray}
and  $\delta f_{q}^{a(i)}$, $\delta f_{\overline{q}}^{a(i)}$, and
$ \delta f_{g}^{a(i)}$  are the partonic fluctuations induced by
the classical field fluctuations
\begin{eqnarray}
\delta f_{q}^{a(i)}(x,\textbf{p}) = -g \int d^{4}y
G_{p}(x-y)p^{\mu} \delta {\cal F}_{\mu \nu}^{a}(y)\frac{\partial
f_{q}(y,\textbf{p})}{\partial p_{\nu}} ,
 \label{flukt-q-sol}\\
\delta f_{\overline{q}}^{a(i)}(x,\textbf{p}) = g \int d^{4}y
G_{p}(x-y)p^{\mu}\delta {\cal F}_{\mu \nu}^{a}(y)\frac{\partial
f_{\overline{q}}(y,\textbf{p})}{\partial p_{\nu}} ,
 \label{flukt-antiq-sol}\\
\delta f_{g}^{a(i)}(x,\textbf{p}) = -g \int d^{4}y
G_{p}(x-y)p^{\mu}\delta {\cal F}_{\mu \nu}^{a}(y)\frac{\partial
f_{g}(y,\textbf{p})}{\partial p_{\nu}} .
 \label{flukt-g-sol}
\end{eqnarray}
Then color current fluctuation can be presented as the sum of
induced current fluctuation, $\delta j^{\mu (i)}(x)$, and
spontaneous current fluctuation, $\delta j^{\mu (s)}(x)$:
\begin{eqnarray}
\delta j^{\mu }_{a}(x)=\delta j^{\mu (i)}_{a}(x)+\delta j^{\mu
(s)}_{a}(x) ,
 \label{current-split}
\end{eqnarray}
where
\begin{eqnarray}
\delta j^{\mu (s)}_{a}(x)=g \int
d^{3}\textbf{p}\frac{p^{\mu}}{2E_{p}} ( \delta
f_{q}^{a(s)}(x,\textbf{p})- \delta
f_{\overline{q}}^{a(s)}(x,\textbf{p})+2
N_{c}\delta f_{g}^{a(s)}(x,\textbf{p})),  \label{current-split-x1} \\
 \delta j^{\mu (i)}_{a}(x)=-g^{2}\int
d^{3}\textbf{p}\frac{p^{\mu}}{2E_{p}} d^{4}y G_{p}(x-y)p^{\alpha}
\delta {\cal F}_{\alpha \beta }^{a}(y)\frac{\partial
f(y,\textbf{p})}{\partial p_{\beta}},
 \label{current-split-x2}
\end{eqnarray}
and
\begin{eqnarray}
f(y,\textbf{p}) \equiv
 f_{q}(y,\textbf{p})+f_{\overline{q}}(y,\textbf{p})+2N_{c}f_{g}(y,\textbf{p})
. \label{funct-all}
\end{eqnarray}
Substituting  Eqs. (\ref{current-split})-(\ref{current-split-x2})
 into the right-hand side of  Eq. (\ref{yang-m-fl}) and  moving the induced current fluctuation into the
 left-hand side,  we get
\begin{eqnarray}
\partial^{\mu}\delta
{\cal F}_{\mu \nu}^{a}(x) + \delta j^{a(i)}_{\nu }(x) = - \delta
j^{a (s)}_{\nu}(x). \label{yang-m-linear}
\end{eqnarray}
 Equation (\ref{yang-m-linear}) can be solved in terms of
four-potential fluctuations, $\delta A_{\mu}^{a}$.  The field
$\delta A_{\mu}^{a}$ is then the sum of the collective field
fluctuations, $\delta A_{\mu }^{a (coll)}$, that are the general
solution of the homogeneous part of Eq. (\ref{yang-m-linear}),
\begin{eqnarray}
\partial^{\mu}\delta
{\cal F}_{\mu \nu}^{a (coll)}(x) + \delta j^{a(i)(coll)}_{\nu }(x)
=0, \label{yang-m-linear-1}
\end{eqnarray}
  and the  field fluctuations $\delta A_{\mu}^{a (part)}$ that are a particular solution of Eq. (\ref{yang-m-linear})
  and are related with particle  fluctuations,
 \begin{eqnarray}
\partial^{\mu}\delta
{\cal F}_{\mu \nu}^{a (part)}(x) + \delta j^{a(i)(part)}_{\nu }(x)
=- \delta j^{a (s)}_{\nu}(x). \label{yang-m-linear-2}
\end{eqnarray}
 Then, because
\begin{eqnarray}
\delta A_{\mu }^{a}(x) = \delta A_{\mu }^{a (coll)} (x)  + \delta
A_{\mu }^{a (part)} (x), \label{funct-coll-all}
\end{eqnarray}
we get
\begin{eqnarray}
\delta f_{q}^{a(i)}(x,\textbf{p})= \delta
f_{q}^{a(i)(coll)}(x,\textbf{p})+ \delta
f_{q}^{a(i)(part)}(x,\textbf{p}),
 \label{current-split-1} \\
 \delta j^{ a(i)}_{\mu}(x) = \delta j^{a
(i)(coll)}_{\mu}(x) + \delta j^{ a (i)(part)}_{\mu}(x).
 \label{current-split-2}
\end{eqnarray}

These formulas can be used  to calculate correlators of
fluctuations ("scattering terms") of averaged over initial
conditions kinetic transport equations of wQGP in the Abelian
approximation. Taking into account traceless of the stress tensor,
$ \delta {\cal F}_{\mu }^{\mu a}(x)=0$, one can rewrite the
right-hand sides of Eqs. (\ref{tr-av-q})-(\ref{tr-av-g}) in a
slightly different form,
\begin{eqnarray}  p^{\mu}
\partial_{\mu} f_{q}(x,\textbf{p}) = - \frac{g}{2N_{c}} \frac{\partial }{\partial p_{\nu}}
 \langle p^{\mu} \delta {\cal
F}_{\mu \nu}^{a}(x)  \delta f_{q}^{a} (x,\textbf{p}) \rangle ,
 \label{tr-av-q-app}\\
 p^{\mu} \partial_{\mu} f_{\overline{q}} (x,\textbf{p}) =
\frac{g}{2N_{c}} \frac{\partial }{\partial p_{\nu}} \langle
p^{\mu} \delta {\cal F}_{\mu \nu}^{a}(x)  \delta
f_{\overline{q}}^{a}(x,\textbf{p}) \rangle ,
 \label{tr-av-antiq-app}\\
 p^{\mu} \partial_{\mu} f_{g}(x,\textbf{p}) = -
\frac{g N_{c}}{ (N^{2}_{c}-1)} \frac{\partial }{\partial p_{\nu}}
\langle  p^{\mu} \delta {\cal F}_{\mu \nu}^{a} (x) \delta
f_{g}^{a} (x,\textbf{p}) \rangle .
 \label{tr-av-g-app}
\end{eqnarray}
Here we take into account that  $t^{a}$, $T^{a}$ matrices satisfy
\begin{eqnarray}
Tr[t^{a}t^{b}]= \frac{1}{2} \delta^{ab}, \label{trace-fund} \\
Tr[T^{a}T^{b}]=N_{c}\delta^{ab}. \label{trace-adj}
\end{eqnarray}
 We present
here  details of the calculation of the scattering term,
$I_{q}^{(Ab)}$,
\begin{eqnarray}
I_{q}^{(Ab)} (x,\textbf{p}) \equiv - \frac{g}{2N_{c}}
\frac{\partial }{\partial p_{\nu}} \langle p^{\mu} \delta {\cal
F}_{\mu \nu}^{a}(x) \delta f_{q}^{a}(x,\textbf{p}) \rangle ,
\label{coll-q-def}
\end{eqnarray}
which governs the evolution of the statistically averaged  quark
distribution function in the Abelian approximation. The
calculation of the other scattering terms is  similar.

 Because of decomposition, Eqs. (\ref{flukt-q-1}) and
(\ref{current-split-1}), one can see that
\begin{eqnarray}
\langle \delta {\cal F}_{\mu \nu}^{a}(x) \delta
f_{q}^{a}(x,\textbf{p}) \rangle = \langle \delta {\cal F}_{\mu
\nu}^{a(coll)}(x) \delta f_{q}^{a(i)(coll)}(x,\textbf{p}) \rangle
+ \nonumber \\  \langle \delta {\cal F}_{\mu \nu}^{a(part)}(x)
\delta f_{q}^{a(i) (part)}(x,\textbf{p}) \rangle + \langle \delta
{\cal F}_{\mu \nu}^{a(part)}(x) \delta f_{q}^{a(s)}(x,\textbf{p})
\rangle . \label{coll-q-def-1-0}
\end{eqnarray}
Let us assume that statistically averaged distribution functions
are slowly varying in space and time, i.e., assume that the rate
of evolution of the fluctuations is large compared to the inverse
of time and length scales on which $f_{q}$, $f_{\overline{q}}$,
$f_{g}$ vary. To explicitly stress this assumption, we  write
space-time coordinates as \textit{index}, e.g.,
$f_{q(x)}(\textbf{p})$ instead of $f_{q}(x,\textbf{p})$ where
necessary. Then, defining the Fourier transformations $a(k)=\int
d^{4}x \exp(i k x) a(x)$ and $a(x)=\int \frac{ d^{4}k}{(2\pi)^{4}}
\exp(-i k x) a (k)$ for a function $a(x)$, we get from Eqs.
(\ref{flukt-q})-(\ref{flukt-g})
 for the $k$ representation of the induced part of the
particle phase-space distribution functions
\begin{eqnarray}
\delta f^{a(i)}_{q (x)}(k,\textbf{p})= g \left [-g^{\mu \nu} +
\frac{k^{\nu}p^{\mu}}{p_{\sigma}k^{\sigma}+i 0}\right ]
\frac{\partial f_{q (x)}(\textbf{p})}{\partial p^{\nu}} \delta
A_{\mu}^{a}(k) ,
\label{delta-q} \\
 \delta f^{a(i)}_{\overline{q}(x)}(k,\textbf{p})= - g \left [-g^{\mu
\nu} + \frac{k^{\nu}p^{\mu}}{p_{\sigma}k^{\sigma}+i 0}\right ]
\frac{\partial f_{\overline{q} (x)}(\textbf{p})}{\partial p^{\nu}}
\delta
A_{\mu}^{a}(k) , \label{delat-antiq} \\
\delta f^{a(i)}_{g (x)}(k,\textbf{p})= g \left [-g^{\mu \nu} +
\frac{k^{\nu}p^{\mu}}{p_{\sigma}k^{\sigma}+i 0}\right ]
\frac{\partial f_{g (x)}(\textbf{p})}{\partial p^{\nu}} \delta
A_{\mu}^{a}(k) , \label{delta-g}
\end{eqnarray}
and the $k$ representation of the induced color current
fluctuation $\delta j^{\mu (i)}$  is
\begin{eqnarray}
\delta j^{\mu (i)}_{(x)}(k)= -g^{2} \int d^{3}\textbf{p}
\frac{p^{\mu}}{2E_{p}}\frac{\partial f_{(x)}(\textbf{p})}{\partial
p_{\lambda}}  \left [g^{\lambda \nu} -
\frac{k^{\lambda}p^{\nu}}{p_{\sigma}k^{\sigma}+i 0}\right ] \delta
A_{\nu}(k) \equiv - \Pi^{\mu \nu}_{(x)} (k) \delta A_{\nu}(k) .
\label{current-fluct-f}
\end{eqnarray}
Using Eq. (\ref{current-fluct-f}), we get the following
expression\footnote{Note that the obtained equation is the same as
that for Abelian plasmas, and if the right-hand side is assumed to
be equal to zero the equation coincides with the linear response
method equation for electromagnetic plasmas and wQGP (see, e.g.,
Refs. \cite{Mrow5,Mrow3,Mrow1,Mrow2,Mrow4}).} for Fourier
transformed Eq. (\ref{yang-m-linear}):
\begin{eqnarray}
(k^{2}g^{\mu \nu}-k^{\mu}k^{\nu}-\Pi^{\mu \nu}_{(x)}(k))\delta
A_{\nu}(k)\equiv \epsilon^{\mu \nu}_{(x)}(k) \delta A_{\nu}(k)
=-\delta j^{\mu (s)}(k) , \label{yang-m-fl-f}
\end{eqnarray}
where the spontaneous color current fluctuation in fundamental
representation, $\delta j^{ (s)}_{\mu }(k)=\delta j^{a(s)}_{\mu
 }(k) t_{a}$,  is Fourier transformed Eq.
(\ref{current-split-x1}):
\begin{eqnarray}
\delta j^{a(s)}_{\mu }(k)=  g \int d^{3}\textbf{p}
\frac{p^{\mu}}{2E_{p}} (\delta f_{q }^{a(s)}(k,\textbf{p})-\delta
f_{\overline{q} }^{a(s)}(k,\textbf{p})+2N_{c}\delta f_{g
}^{a(s)}(k,\textbf{p}) ) . \label{current-f}
\end{eqnarray}
Then the formal solution of   Eq. (\ref{yang-m-fl-f}) as for
four-potential fluctuations,
  $\delta A_{\nu}^{a}(k)$, is
\begin{eqnarray}
\delta A_{\nu}^{a}(k) = \delta A_{\nu}^{a(coll)}(k)+\delta
A_{\nu}^{a(part)}(k),
\label{yang-m-sol1}\\
 \epsilon^{\mu \nu}_{(x)}(k) \delta A_{\nu}^{a(coll)}(k) =0,
 \label{yang-sol1} \\
 \delta A_{\nu }^{a(part)}(k)= - ( \epsilon^{-1} )_{\nu \mu (x)}(k)
\delta j^{\mu a (s)}(k). \label{curr1}
\end{eqnarray}
Here the matrix $( \epsilon^{-1} )_{\nu \mu (x)}(k)$ is the
inverse of $ \epsilon_{\nu \mu (x)}(k)$.

 Now we can  write  the
final expressions for fluctuations of the phase-space densities of
quarks below the nonAbelian scale. Taking into account Eqs.
(\ref{flukt-q-1}), (\ref{yang-m-sol1}), and (\ref{delta-q}), we
get
\begin{eqnarray}
\delta f_{q (x)}^{a}(k,\textbf{p}) = \delta
f_{q}^{a(s)}(k,\textbf{p}) + \delta
f_{q (x)}^{a(i)}(k,\textbf{p})= \nonumber \\
\delta f_{q }^{a(s)}(k,\textbf{p})+ g \left [-g^{\mu \nu} +
\frac{k^{\nu}p^{\mu}}{p_{\sigma}k^{\sigma}+i 0}\right ]
\frac{\partial f_{q (x)}(\textbf{p})}{\partial p^{\nu}} (\delta
A_{\mu }^{a(coll)}(k)+ \delta A_{\mu }^{a(part)}(k)).
\label{delta-q-all-1}
\end{eqnarray}
Then, using  Eq. (\ref{curr1}),  we get
\begin{eqnarray}
\delta f_{q (x)}^{a}(k,\textbf{p}) = \delta f_{q
}^{a(s)}(k,\textbf{p})+ g \left [-g^{\mu \nu} +
\frac{k^{\nu}p^{\mu}}{p_{\sigma}k^{\sigma}+i 0}\right ]
\frac{\partial f_{q (x)}(\textbf{p})}{\partial p^{\nu}} \delta
A_{\mu}^{a(coll)}(k) \nonumber \\
- g \left [-g^{\mu \nu} +
\frac{k^{\nu}p^{\mu}}{p_{\sigma}k^{\sigma}+i 0}\right ]
\frac{\partial f_{q (x)}(\textbf{p})}{\partial p^{\nu}}  (
\epsilon^{-1} )_{\mu \lambda (x)}(k) \delta j^{\lambda
(s)}_{a}(k).
 \label{delta-q-all-2}
\end{eqnarray}

Now, using the relation
\begin{eqnarray}
p^{\mu} \delta {\cal F}_{\mu \nu}^{a}(k)=
(-ip^{\sigma}k_{\sigma}g^{\mu}_{\nu}+ik_{\nu}p^{\mu})\delta
A_{\mu}^{a}(k)  \label{stress-lin}
\end{eqnarray}
and Eq. (\ref{coll-q-def}), we obtain for the scattering term
$I_{q}^{(Ab)}$
 that governs the evolution of the statistically averaged  quark
distribution function the following expression:
\begin{eqnarray}
I_{q}^{(Ab)}(x,\textbf{p}) = -  \frac{g}{2N_{c}}\frac{\partial
}{\partial p_{\nu}}
 \int \frac{
d^{4}k}{(2\pi)^{4}} \frac{ d^{4}k'}{(2\pi)^{4}} \exp(-i (k-k')
x)(-ip^{\sigma}k_{\sigma}g^{\mu}_{\nu}+ik_{\nu}p^{\mu}) \langle
\delta A_{\mu}^{a}(k) f_{q (x)}^{*a}(k',\textbf{p}) \rangle .
 \label{iq-1}
\end{eqnarray}
It is easy to see that $Im[I_{q}]=0$  because $\delta
A_{\mu}^{a}(k)=\delta A_{\mu}^{*a}(-k)$ and $\delta f_{q
(x)}^{a}(k,\textbf{p})=\delta f_{q (x)}^{*a}(-k,\textbf{p})$. Let
us calculate $\langle \delta A_{\mu}^{a}(k) \delta f_{q
(x)}^{a*}(k',\textbf{p}) \rangle$. Taking into account Eqs.
(\ref{yang-m-sol1})-(\ref{curr1}) and (\ref{delta-q-all-2}) we get
\begin{eqnarray}
 \langle \delta A_{\mu}^{a}(k_{1}) \delta f_{q (x)}^{*a}(k_{2},\textbf{p}) \rangle =
 \langle \delta A_{\mu}^{a (coll)}(k_{1}) \delta f_{q (x)}^{*a
(i)(coll)}(k_{2},\textbf{p}) \rangle + \nonumber \\
 \langle \delta A_{\mu}^{a (part)}(k_{1}) \delta f_{q (x)}^{*a
(i)(part)}(k_{2},\textbf{p}) \rangle + \langle \delta A_{\mu}^{a
(part)}(k_{1}) \delta f_{q (x)}^{*a (s)}(k_{2},\textbf{p}) \rangle
,
 \label{iq-1-1}
\end{eqnarray}
where
\begin{eqnarray}
 \langle \delta A_{\mu}^{a (coll)}(k_{1}) \delta f_{q (x)}^{*a
(i)(coll)}(k_{2},\textbf{p}) \rangle = g \left [-g^{\mu^{'} \nu^{
'}} + \frac{k_{2}^{ \nu^{'}}p^{\mu^{'}}}{p_{\sigma}k_{2}^{\sigma}-
i 0}\right ] \frac{\partial f_{q (x)}(\textbf{p})}{\partial
p^{\nu{'}}} \langle \delta A_{\mu(x)}^{a(coll)}(k_{1}) \delta
A_{\mu^{'}(x)}^{*a(coll)}(k_{2}) \rangle , \label{correl-all-1}
\end{eqnarray}
\begin{eqnarray}
\langle \delta A_{\mu}^{a (part)}(k_{1}) \delta f_{q (x)}^{*a
(s)}(k_{2},\textbf{p}) \rangle =  - g \int
d^{3}\textbf{p}'\frac{p'^{\nu}}{2E_{p}'}( \epsilon^{-1} )_{\mu \nu
(x)}(k_{1}) \langle \delta f_{q}^{a(s)}(k_{1},\textbf{p}')\delta
f_{q }^{*a(s)}(k_{2},\textbf{p}) \rangle , \label{correl-all-2}
\end{eqnarray}
\begin{eqnarray}
 \langle \delta A_{\mu}^{a (part)}(k_{1}) \delta f_{q (x)}^{*a
(i)(part)}(k_{2},\textbf{p}) \rangle = g^{3} \int d^{3}\textbf{p}'\frac{p'^{\nu}}{2E_{p}'}  \int d^{3}\textbf{p}''\frac{p''^{\lambda}}{2E_{p}''}\nonumber \\
  ( \epsilon^{-1} )_{\mu
\nu (x)}(k_{1}) \left [-g^{\mu^{'} \nu^{'}} +
\frac{k_{2}^{\nu^{'}}p^{\mu^{'}}}{p_{\sigma}k_{2}^{\sigma}-i
0}\right ] \frac{\partial f_{q (x)}(\textbf{p})}{\partial
p^{\nu^{'}}}  ( \epsilon^{-1} )^{*}_{\mu^{'} \lambda (x)}(k_{2})
\langle \delta f_{q}^{a(s)}(k_{1},\textbf{p}')\delta f_{q
}^{*a(s)}(k_{2},\textbf{p}'') \rangle . \label{correl-all-3}
\end{eqnarray}
Here we take into account that the correlator of the independent
fluctuations is equal to zero.

Now the calculation of the collision term
$I_{q}^{(Ab)}(x,\textbf{p})$ is reduced to finding the explicit
expressions for correlators in Eqs.
(\ref{correl-all-1})-(\ref{correl-all-3}). Let us start with
$\langle \delta A_{\mu(x)}^{a(coll)}(k) \delta
A_{\mu^{'}(x)}^{*a(coll)}(k') \rangle$.  One can easily  get from
 Eq. (\ref{yang-sol1}) (by multiplying the corresponding equation  on $\delta A_{\mu^{'}(x)}^{*a(coll)}(k')$ and
  taking the statistical average) that $\langle \delta A_{\mu(x)}^{a(coll)}(k)
\delta A_{\mu^{'}(x)}^{*a(coll)}(k') \rangle$ is proportional
 to $\delta (| \epsilon^{\mu \nu}_{(x)}(k)|)$
  where $ | \epsilon^{\mu \nu}_{(x)}(k)| $ denotes the determinant of $\epsilon^{\mu \nu}_{(x)}(k)$.
Let us assume that $\langle \delta A_{\mu}^{a(coll)} \delta
A_{\mu^{'}}^{a(coll)} \rangle$ is slowly varying in space-time
function. Then the correlation function of collective classical
field fluctuations is
\begin{eqnarray}
\langle \delta A_{\mu(x)}^{a(coll)}(k) \delta
A_{\mu^{'}(x)}^{*a(coll)}(k') \rangle = (2 \pi)^{4}
\delta^{(4)}(k-k') \delta (| \epsilon^{\mu \nu}_{(x)}(k)|)I_{\mu
\mu^{'} (x)}(k,t). \label{corr-def1}
\end{eqnarray}
Here
\begin{eqnarray}
 I_{\mu \mu^{'} (x)}(k,t)=  I_{\mu \mu^{'} }(\mathbf{k})\exp(2 i k_{0} t ), \label{corr-def2}
\end{eqnarray}
 $I_{\mu \mu^{'} }(\mathbf{k})$ fixes  the strength of
initial correlations of collective field modes and $k_{0} =
\omega_{ (x)}(\mathbf{k}) $  is the turbulent self-energy solution
of the equation
 $ |  \epsilon^{\mu \nu}_{(x)}(k)| = 0 $.

The other correlators in  Eqs.
(\ref{correl-all-1})-(\ref{correl-all-3}) can be expressed through
$\langle \delta f_{q_{i} }^{a(s)}(k,\textbf{p}) \delta f_{q_{j}
}^{*a(s)}(k',\textbf{p}') \rangle $, where $q_{i}$, $q_{j}$  are
$q,\overline{q}$ or $g$. Because  the spontaneous particle phase
density fluctuations are the solution of  Eq. (\ref{flukt-free}),
the corresponding correlation function should be proportional to
$\delta (p^{\mu}k_{\mu})$. The complete lack of initial
correlations in velocities and positions means that the
correlation function should contain
$\delta^{(3)}(\mathbf{p}-\mathbf{p'})\delta^{(4)}(k-k')$. Then,
assuming statistical Poisson fluctuations (and, so, neglecting
quantum statistics, etc.)  and assuming, thereby,  that the
average value of the fluctuation in the squared number of
particles in a certain volume $V$, $\langle \delta N^{2} \rangle$,
is equal to the average number $\langle N \rangle$ of particles in
the volume $V$, we get
\begin{eqnarray}
\langle \delta f_{q_{i} }^{a(s)}(k,\textbf{p}) \delta f_{q_{j}
}^{*a(s)}(k',\textbf{p}') \rangle = (2 \pi)^{8}(2 \pi)^{-3}
\delta_{ij}E_{p} f_{q_{i}}(x,\textbf{p}) \delta^{(4)}(k-k')
\delta^{(3)}(\mathbf{p}-\mathbf{p'}) \delta (p^{\mu}k_{\mu}).
\label{corr-def3-stat}
\end{eqnarray}

Inserting  expressions for correlation functions (\ref{corr-def1})
and  (\ref{corr-def3-stat}) into Eqs.
(\ref{correl-all-1})-(\ref{correl-all-3}) and then  into  Eq.
(\ref{iq-1-1}),
 one can see that
the collision term, $I_{q}^{(Ab)}$, in Eq. (\ref{iq-1}) has the
Fokker-Planck form with transport coefficients of drag, $
A_{\nu(x)}(\textbf{p})$, and diffusion, $ D_{\nu
\mu(x)}(\textbf{p})$:
\begin{eqnarray}
I_{q}^{(Ab)}(x,\textbf{p})=\frac{\partial}{\partial p_{\nu}}\left
( D_{\nu \mu(x)}(\textbf{p})\frac{\partial f_{q
}(x,\textbf{p})}{\partial p_{\mu}}\right ) +
\frac{\partial}{\partial p_{\nu}} \left (
A_{\nu(x)}(\textbf{p})f_{q}(x,\textbf{p})\right ).
 \label{coll-foccer}
\end{eqnarray}
Here
 \begin{eqnarray}
 D_{\nu \mu(x)}(\textbf{p})=D_{\nu \mu(x)}^{(coll)}(\textbf{p}) + D_{\nu
 \mu(x)}^{(part)}(\textbf{p}),
 \label{foccer-split}
\end{eqnarray}
$D_{\nu \mu(x)}^{(coll)}(\textbf{p})$ are the quasilinear
diffusion coefficients that appear due
 to collective field fluctuations, and  $D_{\nu \mu(x)}^{(part)}(\textbf{p})$ are the diffusion
 coefficients  due to particle fluctuations. It is noteworthy that in the quasilinear approximation the
 collective field fluctuations contribute only to the diffusion coefficients of the Fokker-Planck equation
  \cite{Ts}. The
 quasilinear diffusion terms describe the
isotropization processes in momentum space, while the generalized
Balesku-Lenard terms  \cite{Markov} $D_{\nu
 \mu(x)}^{(part)}(\textbf{p})$, $A_{\nu(x)}(\textbf{p})$ lead to thermalization and (local)
equilibrium - the next stage after isotropization.\footnote{It is
noteworthy that the kinetic equation
 with Balescu-Lenard collision terms (as well as with Landau or Boltzmann ones) is, in the thermodynamic sense,
 a kinetic equation of ideal gas: interaction contributes to dissipative quantities and does not
contribute to thermodynamic ones. Then the Balescu-Lenard
collision terms are, streakily speaking,
  improper  near the equilibrium state if the state is far from the ideal gas one.}
In general, the Balescu-Lenard thermalization  time is larger than
the  time scale  of isotropization. Then, for relevant time
scales,
\begin{eqnarray}
I_{q}^{(Ab)}(x,\textbf{p}) \approx \frac{\partial}{\partial
p_{\nu}}\left ( D_{\nu
\nu^{'}(x)}^{(coll)}(\textbf{p})\frac{\partial f_{q
}(x,\textbf{p})}{\partial p_{\nu^{'}}}\right ),
 \label{coll-diff-ql}
\end{eqnarray}
where the quasilinear diffusion coefficients in the Abelian
approximation  are
\begin{eqnarray}
D_{\nu \nu^{'}(x)}^{(coll)}(\textbf{p})=  - \frac{g^2}{2N_{c}}
 \int \frac{
d^{4}k}{(2\pi)^{4}}
 (-ip^{\sigma}k_{\sigma}g^{\mu}_{\nu}+ik_{\nu}p^{\mu})
  \left (-g^{\mu^{'}}_{ \nu^{ '}} + \frac{k_{
\nu^{'}}p^{\mu^{'}}}{p_{\sigma}k^{\sigma}-i 0}\right ) \delta (|
\epsilon^{\mu \nu}_{(x)}(k)|) I_{\mu \mu^{'}(x)}(k,t).
\label{iqd-fin}
\end{eqnarray}

\section{Concluding remarks}
 The isotropization  time  is
the typical time at which the quark-gluon distributions become
locally isotropic in momentum space due to diffusion in momentum
space, and an important question is  under what conditions
isotropization of wQGP can be reached within time scales of
relativistic heavy ion collisions. It is noteworthy that the value
of isotropization time\footnote{Because the corresponding
collision terms depend, in general, on the nonAbelian dynamics of
the gauge fields even at weak coupling, the value of
isotropization time can be calculated only by means of numerical
methods.} depends not only on the instability rate but also on the
initial value of the diffusion coefficient, and, thereby, on the
initial value of the correlation function of collective field
fluctuations.

How the diffusion leads to the spreading  out of the width of the
distribution can be seen from the following simple mathematical
example. Let us consider the diffusion equation
\begin{eqnarray}
 \frac{\partial f(\omega , t)}{\partial t}=D(t) \frac{\partial^{2} f(\omega , t)}{\partial
 \omega^{2}}.
 \label{mod-dif1}
\end{eqnarray}
The solution can be written  as
\begin{eqnarray}
 \int d \omega'G(\omega, \omega', t)f(\omega',t_0),
 \label{mod-dif2}
\end{eqnarray}
where the Green's function
\begin{eqnarray}
 G(\omega, \omega', t)=\frac{1}{(4 \pi d (t))^{1/2}}\exp\left(-\frac{(\omega - \omega')^{2}}{4d(t)}\right)
 \label{mod-dif2}
\end{eqnarray}
represents the function that obeys Eq. (\ref{mod-dif1}) and equals
$\delta (\omega - \omega')$ at $t=t_0$, and
\begin{eqnarray}
d(t)=\stackrel{t}{ \mathrel{\mathop{\int
}\limits_{t_{0}}}}dt'D(t').
 \label{d-D}
\end{eqnarray}
Then the width of $f(\omega , t)$ is determined by the time
dependence of the diffusion coefficient $D(t)$. If, e.g.,
$D(t)=D(t_{0})\exp(\gamma (t-t_{0}) )$, then
$d(t)=\frac{D(t_{0})}{\gamma }(\exp(\gamma (t-t_{0}))-1)$ and for
$\gamma > 0$ increases with time.

Then, keeping in mind relations (\ref{coll-qq-nonab-2}),
(\ref{coll-qq-nonab-4}),  (\ref{corr-def1}), and (\ref{iqd-fin})
between the diffusion term and the correlator of the fluctuations,
we can suppose that, if energy stored in the initial collective
fields' fluctuations is small (say, $D(t_{0})$ is small), then
even for strong initial anisotropy (high values of $\gamma$)
during some initial transient time  the width of the distribution
function will change rather slowly with time ($d(t)$ is small),
and only after growth of collective field modes will  the momentum
width of distribution function start to grow. On the other hand,
if the initial collective fields' fluctuations are strong enough,
then the momentum width of the distribution function can grow
rapidly from the very beginning, and in this case the system can
reach the isotropical state while there are no developed
instabilities  with rapid growth of the fields, as have been noted
in Ref. \cite{{Dum}}.

Note that any initial fluctuations of the color fields are
determined by fluctuations of the sources, and  in the CGC
approach  fluctuations of the sources result from  fluctuations in
the color charge density in each of the colliding nuclei. Also,
violations of boost invariance (really nucleus is not contracted
into an infinitely thin sheet as was assumed in the original
McLerran-Venugopalan model \cite{MV}) result in the strengthening
of fluctuations of the color fields' sources.  As a result of
these perturbations  extremely  disordered color field
configurations \cite{Fuk} can appear. Then, if strong fluctuations
take place, they,  according to our analysis, generate effective
collision terms that  would result in fast local isotropization in
momentum space of statistically averaged particle phase-space
densities.\footnote{ Note that  high initial fluctuations of the
Yang-Mills fields could result not only in  soft modes' (classical
gluon fields) initial fluctuations but also in strengthening of
hard modes' ("particles") initial fluctuations. It was argued
\cite{Kad} that for systems of particles with Coulomb interaction
such "ballistic" modes can lead to the violent relaxation of the
distribution function to a corresponding quasistationary state
(Lynden-Bell distribution \cite{Lynden}) with isotropical velocity
distribution. Such "collisionless" violent relaxation is,
probably, responsible for galaxy dynamics: for most of the
astrophysical systems  the relaxation time for collisional process
exceeds the age of the universe; on the other hand, observations
suggest that for most of the stellar systems velocity distribution
is close to the Maxwellian one \cite{Saslaw}.} A  careful analysis
of the initial conditions after a heavy ion collision would thus
be a key ingredient in understanding the process of
isotropization.

\section*{Acknowledgments}
I am grateful to Yu.M. Sinyukov for discussions. This work was
supported by the Fundamental Research State Fund of Ukraine,
Agreement No. F25/239-2008, and the Bilateral Award DLR (Germany)
- MESU (Ukraine) for the UKR 06/008 Project, Agreement No.
M/26-2008.

\end{document}